\documentclass[a4paper,10pt]{article}
\usepackage{amssymb}
\usepackage{amsmath}
\usepackage{multirow}
\usepackage{graphicx}
\def\d#1{{\rm d}#1}
\def\vp{\varphi}
\DeclareMathOperator{\sech}{sech}

\DeclareMathOperator{\nc}{nc}
\DeclareMathOperator{\F}{F}
\DeclareMathOperator{\E}{E}
\DeclareMathOperator{\K}{K}
\def\sc#1{{\rm sc}#1}

\begin{document}

\title{Globally maximal timelike geodesics in 
static spherically symmetric spacetimes: radial geodesics in static spacetimes and arbitrary 
geodesic curves in ultrastatic spacetimes}
\author{Leszek M. Soko{\l}owski,  Zdzis{\l}aw A. Golda}
\date{}
\maketitle


\begin{center}
{Astronomical Observatory, Jagiellonian University,\\  ul. Orla 171, 30--244 Krak\'ow, Poland}\\
\smallskip\smallskip
{and}\\
\smallskip\smallskip 
{Copernicus Center for Interdisciplinary Studies,\\ ul. S{\l}awkowska 17, 31--016 Krak\'ow, Poland}
\end{center}
\bigskip
\bigskip 
\begin{abstract}
This work deals with intersection points: conjugate points and cut points, of timelike geodesics emanating 
from a common initial point in special spacetimes. The paper contains three results. First, it is shown that 
radial timelike geodesics in static spherically symmetric spacetimes are globally maximal (have no cut points) 
in adequate domains. Second, in one of ultrastatic spherically symmetric spacetimes, Morris--Thorne 
wormhole, it is found which geodesics have cut points (and these must coincide with conjugate points) and 
which ones are 
globally maximal on their entire segments. This result, concerning all timelike geodesics of the wormhole, is the 
core of the work. The third outcome deals with the astonishing feature of all ultrastatic spacetimes: they 
provide a coordinate system which faithfully imitates the dynamical properties of the inertial reference frame. 
We precisely formulate these similarities.
\end{abstract}

\bigskip

\def\hang{\hangindent\parindent}
\def\textindent#1{\indent\llap{#1\enspace}\ignorespaces}
\def\litem{\par\hang\textindent}
\def\subitem{\par\indent \hangindent2\parindent\textindent}
%
%
%
%

\section{Introduction}
\label{sec:01}

This paper is a subsequent work in a sequence of papers on geodesic structure of spacetimes with high symmetries~\cite{S1,SG2,SG1,SG3,SG4,S2}. This research program has developed from the ``twin paradox'' in curved spacetime which has purely geometric nature. It comprises search for longest segments of timelike geodesics and consists in determining intersection points of some geodesics emanating from a common initial point. This in turn comprises two problems: the local one of finding out conjugate points  where close geodesics may intersect and the global problem of determining cut points being the end points of globally longest timelike curves connecting a given pair of points. The global problem is clearly more difficult. In both the problems one needs an exact analytic description of timelike geodesics in terms of known functions and numerical calculations play only a secondary and auxiliary role. This restricts the possible search to spacetimes with a large group of isometries and within them to specific classes of geodesic curves. An exception are ultrastatic spherically symmetric manifolds where all timelike geodesics can be given in explicit form in terms of integrals of metric functions. The static spherically symmetric spacetimes are most investigated in the search for conjugate and cut points on their distinguished 
geodesics, radial and circular. 

In this studying particular spacetimes a question arises: do the same isometries of distinct spacetimes imply 
that their geodesics have the same (or at least similar) structure of conjugate and cut points? On the contrary, 
the example of de Sitter and anti-de Sitter spacetimes shows that this is not the case~\cite{SG2}. The spacetimes 
that have already been analyzed, indicate that no such general rule exists and a set of geodesic intersection 
points may be identified only after making a detailed investigation of some classes of geodesics in the 
given spacetime. Yet it does not mean that one finds a kind of chaos in these spacetimes and two examples of common 
geodesic structure are known in static spherically symmetric manifolds: 
the radial timelike  geodesics are globally maximal and the geodesic deviation equation on circular geodesics (if they exist) is the same in each of these spacetimes~\cite{SG3}.

The paper is organized as follows. In Section 2 we give a complete and detailed proof of the proposition that 
the radial timelike geodesics in static spherically symmetric spacetimes are globally maximal on their segments 
lying in the domain of the chart explicitly exhibiting these isometries. The idea of the proof previously 
appeared in the printed version of~\cite{SG3} whereas the arXiv version of that work contains an incomplete 
proof. Section 3 discusses cut points and conjugate points of any timelike geodesic in a particular 
ultrastatic spherically symmetric spacetime: Morris--Thorne wormhole; this is the main result of this work. 
It turns out that spacetimes even in this very narrow class of manifolds may show some differences in this 
aspect. It has been known that all ultrastatic spacetimes (also without spherical symmetry) expressed in comoving coordinates may imitate the inertial reference frame. These similarities have usually been presented in an incomplete an imprecise way~\cite{SS}. For this reason we give in Appendix a proposition stating to what extent the comoving system in these spacetimes imitates (i.e. has the properties of) the inertial frame. Throughout the paper we are dealing exclusively with timelike geodesics and this is not always marked.

\section{Globally maximal segments of radial timelike geodesics in static spherically symmetric spacetimes}
\label{sec:02}

In this section we consider static spherically symmetric (SSS) spacetimes; in these manifolds radial timelike geodesics are distinguished by their geometric simplicity and physical relevance. Among all coordinate systems which are adapted to spherical symmetry we single out the standard ones which make the metric explicitly time independent and diagonal, what implies that the timelike Killing vector field is orthogonal to constant time spaces. Radial geodesics in these coordinates are those with their tangent vector spanned on vectors tangent to coordinate lines of time and radial variable. The metric of any SSS spacetime in the standard coordinates is
	\begin{equation}
\d{s}^2=e^{\nu(r)}\d{t}^2-e^{\lambda(r)}\d{r}^2-F^2(r)
\left(
\d{\theta}^2+\sin^2\theta \d{\vp}{}^2
\right),
\label{eq:metric}
	\end{equation}
where $t\in(-\infty,+\infty)$, arbitrary functions $\nu$ and $\lambda$ are real for $r\in(r_m,r_M)$ and $t$ and $r$ have dimension of length. $F(r)=r$ for a generic 
SSS metric and 
$F(r)=\mbox{const}$ for some specific spacetimes (e.g. Bertotti--Robinson metric). 
To avoid inessential complications we assume that the spacetime is asymptotically flat at spatial infinity, i.e. $\nu$ and $\lambda$ tend to zero for $r\to r_M\leq\infty$. Then the timelike Killing vector is normalized at spatial infinity and is $K^\alpha=\delta_0^\alpha$. Let a timelike geodesic be the worldline of a particle of mass $m$, then the conserved energy is $E=c\, K^\alpha p_\alpha=mc^2\, K^\alpha \dot{x}_\alpha$ and energy per unit rest mass is $k\equiv \frac{E}{mc^2}>0$ and is dimensionless. For the metric~(\ref{eq:metric}) one finds
	\begin{equation}
\dot{t}\equiv\frac{\d{t}}{\d{s}}=k e^{-\nu}.
\label{eq:geodesict}
	\end{equation}
In the rest this section we shall only consider radial timelike geodesics, $t=t(s)$, $r=r(s)$, $\theta=\mbox{const}$, $\varphi=\mbox{const}$ and we assume that they can be extended to infinity going outwards from any point $r_0>r_m$. The universal integral of motion, $g_{\alpha\beta}\dot{x}^\alpha\dot{x}^\beta=1$, allows one to replace the radial component of the geodesic equation by a first order expression
	\begin{equation}
\dot{r}\equiv\frac{\d{r}}{\d{s}}=
\left[
e^{-(\nu+\lambda)}
\left(
k^2
-e^\nu
\right)
\right]^{\frac12}.
\label{eq:geodesicr}
	\end{equation}
Since $\nu\to0$ at spatial infinity, one gets $k\geq1$ and the geodesic may be extended inwards for all $r$ such that $k^2\geq e^\nu$. For simplicity we assume that $e^\nu$ is bounded in the chart domain and $k^2\geq e^\nu$ everywhere in it. 

To show that radial geodesics are globally maximal one considers a three-dimensional congruence of radial 
geodesics with the same energy $k$ filling the entire chart domain. The congruence is described by the 
geodesic velocity field 
	\begin{equation}
u^\alpha=\frac{\d{x^\alpha}}{\d{s}}=(\dot{t},\dot{r},0,0)=
\left[
k e^{-\nu}, e^{-\frac12 (\nu+\lambda)}
\left(
k^2
-e^\nu
\right)^\frac12,0,0
\right].
\label{eq:geodesictr}
	\end{equation}
We shall use this vector field to construct the comoving (Gauss normal geodesic, GNG) coordinate system adapted 
to this congruence. First, the field is rotation free, $\nabla_{[\alpha}u_{\beta]}$=0, hence it is a gradient field, $u_\alpha=\partial_\alpha\tau$ where $\tau=\tau(t,r)$ and the congruence is orthogonal to hypersurfaces $\tau=\mbox{const}$. One easily finds $\tau$ from equations
	\begin{equation}
\frac{\partial \tau}{\partial t}=u_0=k~~\mbox{and}~~\frac{\partial \tau}{\partial r}=u_1=
- e^{\frac12 (\lambda-\nu)}\left(k^2
-e^\nu
\right)^\frac12,
\label{eq:geodesicp}
	\end{equation}
then
	\begin{equation}
\tau=kt-\int\!\!
\left[
e^{\lambda-\nu}\left(k^2
-e^\nu
\right)
\right]^\frac12\d{r}.
\label{eq:geodesictau}
	\end{equation}
$\tau$ will be used as a new time coordinate. A new coordinate system is completed by introducing a new radial variable $R$ whose coordinate lines must be orthogonal to $\tau$ lines. One postulates $R(t,r)=at+P(r)$, where $a=\mbox{const}$ and $P(r)$ will be determined by the orthogonality condition. The relationship for the differentials $\d{\tau}=k\d{t}+u_1\d{r}$ and $\d{R}=a\d{t}+P'\d{r}$, $P'=\frac{\d{P}}{\d{r}}$, is inverted to 
	\begin{equation}
\d{r}=\frac{1}{H}
\left(
k\d{R}-a\d{\tau}
\right),~~\d{t}=\frac{1}{H}
\left(
P'\d{\tau}-u_1\d{R}
\right),
\label{eq:geodesicrt}
	\end{equation}
where $H(r)\equiv k P'-a u_1$. Inserting $\d{t}$ and $\d{r}$ into the line element~(\ref{eq:metric}) one finds that the coefficient at $2\d{\tau}\d{R}$ is 
$g_{01}'=-e^\nu P' u_1+a k e^\lambda$ and requiring $g_{01}'=0$ one gets $P'=-a k\left[e^{\lambda-\nu}(k^2-e^\nu)^{-1}\right]^\frac{1}{2}$. It turns out that the 
factor $a$ is redundant and one puts $a=1$, then the coordinate transformation is (\ref{eq:geodesictau}) and
	\begin{equation}
R=t-k\int \!\!
\left[e^{\lambda-\nu}(k^2-e^\nu)^{-1}\right]^\frac{1}{2}\d{r}=t+P(r).
\label{eq:geodesicR}
	\end{equation}
One finds the inverse transformation first by determining $r=r(\tau,R)$. The latter arises from the difference
	\begin{equation}
\tau-k R=\int\!\! 
\left[e^{\lambda-\nu}(k^2-e^\nu)^{-1}\right]^\frac{1}{2}\d{r}\equiv W(r,k).
\label{eq:geodesictauR}
	\end{equation}
Function $W$ is positive and monotonically growing since $\frac{\d{W}}{\d{r}}>0$. Hence $W$ is invertible in the whole range $r_m<r<r_M$. One sees that $-\infty<\tau<+\infty$ and $-\infty<R<+\infty$, however the chart domain does not cover the whole $(\tau,R)$ plane. Actually $\tau$ and $R$ vary in the strip
	\begin{equation}
W(r_m,k)<\tau-k R<W(r_M,k).
\label{eq:geodesictauW}
	\end{equation}
Next $r=W^{-1}(\tau-k R)$ and inserting it into (\ref{eq:geodesicR}) one finds $t=R-P[W^{-1}(\tau-k R)]$. 
As an example we take Reissner--Nordstr\"om black hole, $M^2>Q^2$. For it
	\begin{equation}
e^\nu=1-\frac{2M}{r}+\frac{Q^2}{r^2}=e^{-\lambda},
\label{eq:geodesicenua}
	\end{equation}
here $r\in (r_+,\infty)$ with $r_+=M+\sqrt{M^2-Q^2}$. We do not consider the maximally extended spacetime and assume existence of only one exterior asymptotically flat region. Outside the outer event horizon there is $e^\nu<1$ and one sets $k=1$ for simplicity, then
	\begin{equation}
W(r,1)=\frac23(2M)^{-\frac12}\sqrt{r-\frac{Q^2}{2M}}\left(r+\frac{Q^2}{M}\right).
\label{eq:geodesicenub}
	\end{equation}
$W$ grows from $W(r_+)=\frac{r_+}{3M}\left(r_+ + \frac{Q^2}{M}\right)$ to infinity for $r\to\infty$, yet it cannot be effectively inverted since it requires solving a 
cubic equation. Only for $Q=0$ one gets $r=(2M)^\frac13\left[\frac23(\tau-R)\right]^\frac23$ and an explicit expression for $t=t(\tau, R)$ may be found 
(Eddington--Lema\^{\i}tre coordinates). For Kottler (Schwarzschild--de Sitter) black hole for $0<\Lambda<\frac{1}{9M^2}$ the 
corresponding formulae are quite complicated.\\
One sees from the construction that for arbitrary $\nu$ and $\lambda$ the domains for $(t,r)$ and $(\tau,R)$ charts form the same region of the spacetime.

The transformation (\ref{eq:geodesictau}) and (\ref{eq:geodesicR}) provides a specific comoving system and the 
SSS metric (\ref{eq:metric}) reads in it
	\begin{equation}
\d{s}^2=\d{\tau}^2-
\left[k^2-e^\nu
\right]\d{R}^2 -F^2\left[
r(\tau{\mathit-R})
\right]\d{\mathit\Omega}^2,
\label{eq:metrictauR}
	\end{equation}
where as usual, $\d{\mathit\Omega}{}^2=\d{\theta}{}^2+\sin^2\theta \d{\vp}{}^2$. The metric does not explicitly depend on $\lambda(r)$ since this function has been 
absorbed in the transformation, however the metric depends implicitly on $\lambda$ and explicitly on time $\tau$ via the inverse transformation $r=W^{-1}(\tau-R)$. 
For special spacetimes (Schwarzschild, Reisner--Nordstr\"om) the metric (\ref{eq:metrictauR}) may be, as is well known, extended to a domain larger than that in 
$(t,r)$ chart. 

Next, to show that the GNG system of $(\tau,R)$ coordinates is actually a comoving one in the sense that 
the congruence of radial timelike geodesics generating it is a family of coordinate lines for time $\tau$ (the particles are at rest), one transforms the velocity field $u^\alpha$ given in (\ref{eq:geodesictr}) to this 
system. One gets $\frac{\d{\tau}}{\d{s}}=1$ and $\frac{\d{R}}{\d{s}}=0$. Each radial geodesic is described 
by $\tau=\tau_0+s$, $R=R_0$, $\theta=\theta_0$ and $\vp{}=\vp{}_0$. 

Now the proof that each radial geodesic extended to the whole range from $r_m$ to $r_M$ is globally maximal between any pair of its points, is immediate. Let 
$P_0(\tau_0,R_0,\theta_0,\vp{}_0)$ and $P_1(\tau_1,R_0,\theta_0,\vp{}_0)$ lie in the strip (\ref{eq:geodesictauW}), then they are connected by a unique radial 
geodesic $\gamma$. Its length is $s(\gamma)=\tau_1-\tau_0$. For any timelike curve $\sigma$ connecting $P_0$ and $P_1$ and parametrized by 
$\tau$, $x^\mu=x^\mu(\tau)$, its length is 
	\begin{equation}
s(\sigma) = \int\limits_{\tau_0} ^{\tau_1}
\left[
1-(k^2-e^\nu)\left(\frac{\d{R}}{\d{\tau}}\right)^2-
F^2\left(\frac{\d{\theta}}{\d{\tau}}
\right)^2
-F^2\sin^2\theta\left(\frac{\d\vp{}}{\d{\tau}}\right)^2
\right]^\frac12\d{\tau}<\tau_1-\tau_0,\nonumber 
\label{eq:ssigma}
	\end{equation}
hence $\gamma$ is globally the longest curve.

\section{Future cut points on arbitrary timelike geodesics and globally maximal curves in Morris--Thorne 
wormhole} 
\label{sec:03}

In most SSS spacetimes the geodesic equation for timelike curves cannot be effectively integrated (to give a parametric description) besides special cases such as radial 
and circular lines, hence finding the longest curve joining two arbitrary (chronologically related) points is a hopeless task. Yet explicit formulae for arbitrary 
timelike geodesics in terms of integrals have been found for ultrastatic spherically symmetric (USSS) spacetimes. Generic ultrastatic spacetimes (without spherical 
symmetry) are described in~\cite{SS} and arbitrary timelike geodesics in any USSS spacetime are given in~\cite{SG1}. Even in the USSS case these integral formulae 
for parametric description of geodesic lines do not allow one to determine whether a given curve is globally maximal on its sufficiently long segment. One must 
instead separately study particular USSS spacetimes in which any timelike geodesic is explicitly expressed in terms of known functions. In this section we investigate 
the geodesic structure of the Morris--Thorne wormhole. Its properties are discussed in~\cite{MT} and for our purpose we only need its metric expressed in a chart 
covering the entire manifold. In~\cite{MT} it was defined as a special case of USSS spacetimes with the metric
	\begin{equation}
\d{s}^2=\d{t}^2-\frac{r^2}{r^2-a^2}\d{r}^2-r^2\d{\mathit\Omega}^2,
\label{eq:metricUSSSa}
	\end{equation}
where $a=\mbox{const}>0$, i.e. it is metric~(\ref{eq:metric}) with $\nu=0$ and $F(r)=r$. However this chart has 
a boundary $r=a$ which is a coordinate singularity and the spacetime can be extended beyond it. To this end 
one computes the length of the radial line $t=\mbox{const}$ from the boundary $r=a$ to a point $r=r_0$, 
$l(a,r_0)=\int_a^{r_0}(-\d{s}^2)^\frac12=\sqrt{r_0^2-a^2}$. 

One then introduces a new radial coordinate $l=\sqrt{r^2-a^2}$ and the metric is~\cite{MT}
	\begin{equation}
\d{s}^2=\d{t}^2-\d{l}^2-(l^2+a^2)(\d{\theta}^2+\sin^2\theta\,\d\vp{}^2)
\label{eq:metricUSSS}
	\end{equation}
and the singularity disappears. Here $-\infty<t<\infty$ and $\theta$ and $\vp$ cover (in the usual sense) $S^2$. The original domain $a<r<\infty$ or $0<l<\infty$ is 
now extended to entire real line, $-\infty <l <\infty$. The entire manifold has two flat spatial infinites, $l\to\pm\infty$. The former singularity is actually a 
regular hypersurface $l=0$, the ``wormhole throat''. The metric (\ref{eq:metricUSSS}) may be written as $ds^2=a^2 d\bar{s}^2$ and the rescaled metric $d\bar{s}^2$ with 
dimensionless coordinates $t/a$ and $l/a$ has the same geodesic structure as $ds^2$. The rescaled metric is parameter-free, hence properties of geodesic curves on 
the wormhole manifold are independent of the value $a$.                                                   

For a timelike geodesic motion the integral of energy is from (\ref{eq:geodesict}) $\dot{t}=k$ and may be 
integrated to $t(s)=t_0+ks$. The motion is ``flat'' in the sense that each geodesic lies in a 2-plane which 
in adapted to it angular coordinates is given by $\theta=\frac{\pi}{2}$ and the conserved angular 
momentum generated by the Killing field $K_\varphi^\alpha=\delta_3^\alpha$ is 
$g_{\alpha\beta}K_{\varphi}^\alpha p^\beta\equiv-mL$; $L$ is the angular momentum per unit mass and
	\begin{equation}
\dot\varphi\equiv \frac{\d\varphi}{\d s}=\frac{L}{l^2+a^2},
\label{eq:dotvarphi}
	\end{equation}
$L$ has dimensions of length. The geodesic equation for $l$ is replaced by a modified version of 
(\ref{eq:geodesicr}), which reads
	\begin{equation}
\dot {l}^2=k^2-1-\frac{L^2}{l^2+a^2}.
	\label{eq:dotl}
	\end{equation}
For $k=1$ one gets $L^2=\dot{l}^2=\dot{\varphi}=0$, the motion is reduced to a rest (see Appendix) and the 
geodesic is a time coordinate line which is globally maximal. We are interested in all other geodesic curves, hence $k>1$. In most formulae below we shall use a dimensionless radial coordinate $x=\frac{l}{a}$.

Before searching for cut points on arbitrary nonradial geodesics we deal with the following problem: is it possible to connect two points on a radial geodesic (points with the same angular coordinates) by another timelike geodesic?

\subsection{A radial geodesic intersected twice by a nonradial geodesic}

Let $C$ be a radial geodesic with energy $k_C$, $\theta=\frac{\pi}{2}$, $\varphi=\varphi_0$. From (\ref{eq:dotl}) one gets $\dot l=\pm\sqrt{k_C^2-1}$ and one sees that inversion in time, $t\to-t$, maps a radial geodesic with $\dot l=+\sqrt{k_C^2-1}$ onto the geodesic with $\dot l=-\sqrt{k_C^2-1}$. Hence it is sufficient to consider the segment of $C$ with $\dot l>0$ and $l>0$. Choose a point $P_0(t_0,l_0)$ for $l_0>0$ on $C$, then $t(s)=t_0+k_C s$ and $l(s)=\sqrt{k_C^2-1}s+l_0$.

Consider now a nonradial geodesic $G$ which emanates from $P_0$ and goes to $l\to\infty$. We restrict it by requiring $l_0$ be the lowest value of $l(s)$ on $G$. If $l$ attains minimum for $l=l_0$, then $\dot l(l_0)=0$ and~(\ref{eq:dotl}) implies $k_G^2-1-\frac{L^2}{l_0+a^2}=0$.

In general a nonradial geodesic depends on two parameters, $k_G$ and $L$. The above restriction shows that $L^2$ is determined by $k_G$ and $l_0$,
	\begin{equation}
L^2=(k_G^2-1)(l_0^2+a^2)
\label{eq:L2l}
	\end{equation}
and one has a one-parameter family of geodesics $G(k_G)$ emanating from $P_0$. (One may check that the acceleration $\ddot l>0$ on $G$ at $P_0$). To simplify expressions below one introduces parameter
	\begin{equation}
p^2\equiv \frac{k_G^2-1}{L^2}=\frac{1}{l_0^2+a^2},~~\mbox{hence}~~p^2 a^2= \frac{a^2}{l_0^2+a^2}<1.
	\label{eq:p2l}
	\end{equation}
In terms of $x=\frac{l}{a}$ and $x_0=\frac{l_0}{a}$ eq.~(\ref{eq:dotl}) reads for $G$ 
	\begin{equation}
\dot{l}^2=(k_G^2-1)\frac{x^2-x_0^2}{x^2+1}
	\label{eq:dotl2}
	\end{equation}
and its length from $P_0$ is 
	\begin{equation}
s(l)=a(k_G^2-1)^{-\frac{1}{2}}\int\limits_{x_0}^{x}
\left[
\frac{x^2+1}{x^2-x_0^2}\right]^\frac12 \d x\equiv a(k_G^2-1)^{-\frac12}H_s(x_0,x),
	\label{eq:sl}
	\end{equation}
where 
	\begin{equation}
H_s(x_0,x)=x^{-1}\sqrt{\left(x^2+1\right)
   \left(x^2-x_0^2\right)}+\sqrt{x_0^2+1}
   \left[\F\left[\arccos
   \left[\frac{x_0}{x}\right],\frac{1}{x_0^2+1}\right] - \E\left[\arccos
   \left[\frac{x_0}{x}\right],\frac{1}{x_0^2+1}\right]\right]\nonumber
	\end{equation}
and $\F$ is the incomplete elliptic integral of the first kind, and $\E$ is incomplete elliptic integral of the second kind, see~\cite{BF,GR}. Also the angle $\varphi$ 
on $G$ is parametrized by $l$. From (\ref{eq:dotvarphi}) and (\ref{eq:dotl2}) one gets assuming $L>0$,
	\begin{equation}
\varphi(l)=\varphi_0+\sqrt{x_0^2+1}\int\limits_{x_0}^{x}
\left[
({x^2+1})({x^2-x_0^2})
\right]^{-\frac12}\d x\equiv \varphi_0+\sqrt{x_0^2+1}H_\varphi(x_0,x),
	\label{eq:varphil}
	\end{equation}
where 
	\begin{equation}
H_\vp(x_0,x)=\F\left[\arccos
   \left[\frac{x_0}{x}\right],\frac{1}{x_0^2+1}\right].\nonumber
	\end{equation}
The increase of the angle $\vp$ on $G(k_G)$ is independent of energy $k_G$ and is the same for all geodesics of this family.

The necessary condition for $G$ to intersect the radial $C$ at some point $P_1(t_1, l_1)$ is that the increment of $\vp$ on $G$ is $2\pi$,
	\begin{equation}
\Delta\varphi\equiv \varphi(l_1)-\varphi_0=\sqrt{x_0^2+1}H_\varphi(x_0,x_1)=2\pi.
	\label{eq:Deltavarphil}
	\end{equation}
This is a transcendental equation for $x_1=\frac{l_1}{a}<\infty$ and has an analytic solution in the form
 $x_1=x_0 \nc\left(2\pi,\frac{1}{1+x_0}\right)$, where $\nc$ is one of Jacobi elliptic functions \cite{BF,GR}. 
 Rather surprisingly, it turns out that solutions do not exist for arbitrary values of $x_0$ and they exist in a narrow interval $0<x_0<x_{0M}=0,0074705{\ldots}$ For $x_0\to x_{0M}$ one finds $x_1\to \infty$. The lower limit 
$x_0=0$ is unattainable since it is seen from (\ref{eq:varphil}) that $\frac{\d\varphi}{\d l}$ is divergent there 
as $\frac{1}{x}$ and $\vp(l)$ behaves as $-\ln x$ (actually for $x_0=0$ the integral (\ref{eq:varphil}) is an elementary function which is logarithmically divergent in the lower limit $x_0\to 0$), hence $G$ winds up 
 infinitely many times around $l=0$ for infinitesimal $l$. For example one takes two extreme cases:
\litem{--} for $(1+x_0)^{-1}=0,99999$ corresponding to $x_0\cong 0,003162$ there is  $x_1\cong 1,0315483397{\ldots}$,
\litem{--} for $x_0\cong 0{,}00747$ one gets  $x_1=14\,098{,}2$.

\noindent 
The existence of finite solutions $x_1$ to eq. (\ref{eq:Deltavarphil}) raises the question of whether it is possible for $G$ to intersect $C$ more than once. If $n$ intersection points exist, $x_1<x_2<\ldots<x_n$, then 
each of them is a solution to equation analogous to (\ref{eq:Deltavarphil}), 
	\begin{equation}
\Delta\varphi=\sqrt{x_0^2+1}H_\varphi(x_0,x_n)=2\pi n.
	\label{eq:Deltavarphiln}
	\end{equation}
This equation is analytically solved for any natural $n$ by $x_n=x_0 \nc\left(2\pi n,\frac{1}{1+x_0}\right)$.  
The solutions $x_n$ exist for decreasing ranges of initial values of $x_0$. The value 
$x_{0M}\equiv x_{0M_1}=0{,}0074705{\ldots}$ was found numerically, yet there are analytic arguments that it 
may be well approximated by $x_{0M_1}=2\sech(2\pi)$. In the limit $n\to \infty$ one gets an exact expression for the upper limit of the interval $0<x_n<x_{0M_n}$,
	\begin{equation}
x_{0M_n}=2 e^{-2\pi(n-1)}\sech(2\pi)
	\label{eq:x0Mn}
	\end{equation}
and the interval length exponentially diminishes. 

If the necessary condition holds, the sufficient condition for $G$ to intersect $C$ at $P_1$ is that the 
time coordinates of both the curves are equal at this point, 
$t_1=t_0+k_C s_C=t_0+k_G s_G$, where their lengths are respectively 
	\begin{equation}
s_C=a(k_C^2-1)^{-\frac{1}{2}}(x_1-x_0)
	\label{eq:sCq}
	\end{equation}
and $s_G=a(k_G^2-1)^{-\frac{1}{2}}H_s(x_0,x_1)$ from (\ref{eq:sl}). Hence the sufficient condition takes the form 
	\begin{equation}
k_C(k_C^2-1)^{-\frac{1}{2}}(x_1-x_0)=k_G(k_G^2-1)^{-\frac{1}{2}}H_s(x_0,x_1).
	\label{eq:sCsG}
	\end{equation}
For a given energy $k_C$ on $C$ and known solution $x_1(x_0)$ of eq.~(\ref{eq:Deltavarphil}), this is an equation 
for energy $k_G$ on $G$ and this means that $C$ may be intersected at $P_1$ by only one geodesic out of the 
family $G(k_G)$. The solution of (\ref{eq:sCsG}) is 
	\begin{equation}
k_G^2=\frac{k_C^2(x_1-x_0)^2}{k_C^2(x_1-x_0)^2-(k_C^2-1)H_s^2(x_0,x_1)}.
	\label{eq:kG2}
	\end{equation}
This formula makes sense if its denominator is positive, then $k_G^2>1$. The requirement that the denominator 
be positive is in turn a restriction imposed on $k_C$. 
Since $x_1=x_1(x_0)$, one introduces a function
	\begin{equation}
B(x_0)=\frac{H_s^2(x_0,x_1)}{(x_1-x_0)^2}>0
	\label{eq:kG2a}
	\end{equation}
and the requirement takes on the form of an inequality,
	\begin{equation}
k_C^2(B(x_0)-1)<B(x_0).
	\label{eq:kCB}
	\end{equation}
A numerical computation applying Mathematica shows that in the allowed interval $0<x_0<0{,}0074705{\ldots}$ 
function $B$ is diminishing and everywhere $B(x_0)>1$. Then for given $x_0$ the energy $k_C$ is restricted to 
the interval
	\begin{equation}
1<k_C^2<\frac{B(x_0)}{B(x_0)-1}. 
	\label{eq:kG2b}
	\end{equation}
For an admissible value of $k_C$ the length of $G$ which emanates from $P_0$ and intersects $C$ at $P_1$ is, 
from (\ref{eq:sl}) and (\ref{eq:kG2}), 
	\begin{equation}
s_G=a\frac{\sqrt{B-k_C^2(B-1)}}{\sqrt{(k_C^2-1)B}}H_s(x_0,x_1). 
	\label{eq:SG}
	\end{equation}
The ratio of the geodesic lengths is
	\begin{equation}
\left[
\frac{s_G}{s_C}
\right]^2
=B-(B-1)k_C^2. 
	\label{eq:sGsC}
	\end{equation}
and it is clear that $s_G<s_C$ since $B(x_0)>1$ and $k_C^2$ is in the allowed range. We note that this is 
not another proof of geodesic $C$ being globally maximal because 
$G$ is not the most generic nonradial geodesic intersecting $C$ at $P_0$ and $P_1$: $G$ cannot be extended for 
$l<l_0$ since its angular momentum is restricted by (\ref{eq:L2l}).

\subsection{Future cut points on nonradial timelike geodesics}

Now we seek for globally maximal (globally longest) segments of nonradial geodesics in the wormhole spacetime. 
To this end we briefly remind the necessary notions of Lorentzian geometry. The {\it Lorentzian distance function\/} 
$d(p,q)$ of two chronologically related points $p$ and $q$ ($q$ is in the chronological future of $p$, $p\prec\!\prec q$) is the length of the longest timelike curve 
joining $p$ and $q$. The curve $G$ from $p$ to $q$ is said to be {\it globally maximal\/} if it is the longest 
one between these points, i.e. if $s(G)=d(p,q)$. The 
globally maximal curve (usually non unique) is always a timelike geodesic (Theorem 4.13 of~\cite{BEE}).

We consider complete timelike geodesics: they are defined for all values of the canonical length parameter, 
$-\infty<s<+\infty$; in Morris--Thorne wormhole they extend to $l\to\pm\infty$. Usually they are not globally 
maximal beyond some segment, in our notation: from $P_0$ to $P_1$. This gives rise to the notion of the cut 
point on a geodesic. Let $G$ be a future directed timelike geodesic parametrized by its length $s$ and let 
$P_0=G(0)$ be a chosen point on it. Set
	\begin{equation}
s_0\equiv
\sup
\left\{s>0{:}~d
\left(G(0), G(s)
\right)=s
\right\}.\nonumber
	\end{equation}
If $0<s_0<\infty$ then $G(s_0)$ is said to be the {\it future timelike cut point\/} of $G(0)$ along $G$. For all 
$0<s<s_0$ the geodesic $G$ is the unique globally maximal curve from $G(0)$ to $G(s)$ and is globally maximal 
(not necessarily unique) on the segment from $G(0)$ to $G(s_0)$. For $s_1>s_0$ there exists a future 
directed timelike curve $K$ from $G(0)$ to $G(s_1)$ which is longer, $s(K)>s(G)$. In other terms: $s_0$ is the length of the longest globally maximal segment of $G$ 
from $P_0=G(0)$. 

We therefore consider the full set of complete timelike geodesics $G(k,L)$ intersecting at some point $P_0$ 
and seek for another point $P_1$ where some of them intersect again. 
Due to spherical symmetry each geodesic lies in space in its own ``plane'' $\theta=\frac{\pi}{2}$. If two 
geodesics lying in different planes intersect twice, the difference in the azimuthal angle $\varphi$ 
between intersection points (measured in one of the planes) is $\Delta\varphi=\pi$. This geometric argument 
will be analytically shown below. We therefore focus our attention on double intersections of geodesics 
belonging to the same 2-plane. Then the intersection points are $P_0(t_0,l_0,\frac{\pi}{2},\vp)$ and  
$P_1(t_1,l_1,\frac{\pi}{2},\bar\vp)$. For each complete geodesic eq.~(\ref{eq:dotl}) holds 
for all values of $l$ and the minimal value of ${\dot{l}}^2$ is attained for $l=0$ and at this point there must 
be $\dot{l}^2=k^2-1-\frac{L^2}{a^2}>0$, what implies $\left(k^2-1\right)\frac{a^2}{L^2}>1$.

In fact, if $\dot{l}^2(l=0)=0$, then $\left(k^2-1\right)\frac{a^2}{L^2}=1$ and one gets from the geodesic 
equation that $\ddot{l}=0$ at this point and the unique solution of this equation is $l(s)=0$ while 
$\dot{\vp}=\frac{L}{a^2}$. This is a circular geodesic at the wormhole throat. As a side remark we note that 
timelike circular geodesics with different $L$ exist only at the throat, $l=0$. As in (\ref{eq:p2l}) we denote 
$p^2=(k^2-1)L^{-2}$, hence $p^2a^2>1$. We assume that $\d l>0$ for $\d s>0$ and $-\infty <L<\infty$. In terms of 
$p^2$ and $x=\frac{l}{a}$ one finds from (\ref{eq:dotl})
	\begin{equation}
\frac{\d l}{\d s}=\frac{|L|}{a}
\left(
x^2+1
\right)^{-\frac12}
\left[
p^2a^2(x^2+1)-1
\right]^\frac12.
	\label{eq:dlds}
	\end{equation}
This expression may be given a more convenient form if one introduces a parameter
	\begin{equation}
b^2\equiv 1-\frac{1}{p^2a^2}=1-\frac{L^2}{(k^2-1)a^2},~~0<b^2<1.
	\label{eq:b2}
	\end{equation}
Integrating the inverse of formula (\ref{eq:dlds}) one gets the length of a generic nonradial geodesic 
$G(k,L)$ from $l_0$ to $l$, 
	\begin{equation}
s(k,L,l)=\frac{a}{\sqrt{k^2-1}}\int\limits_{x_0}^x
\left[
\frac{x^2+1}{x^2+b^2}
\right]^\frac12\d x\equiv \frac{a}{\sqrt{k^2-1}}J(b,x_0,x).
	\label{eq:skLl}
	\end{equation}
Analogously to derivation of (\ref{eq:varphil}) one has in the general case from (\ref{eq:dotvarphi}) and 
(\ref{eq:dlds}) that the angular coordinate along $G$ is 
	\begin{equation}
\vp(l_0,l)=\vp_0+\frac{L}{|L|}\sqrt{1-b^2}\int\limits_{x_0}^x
\left[
(x^2+1)(x^2+b^2)
\right]^{-\frac{1}{2}}\d x\equiv \vp_0+\frac{L}{|L|}\sqrt{1-b^2}N(b,x_0,x).
	\label{eq:vpl0la}
	\end{equation}
Assume that two geodesics of the set, $G_1(k_1,L_1)$ and $G_2(k_2,L_2)$, intersect at~$P_1$. First, one finds 
that $G_1$ and $G_2$ emanating from $P_0$ cannot intersect again if their angular momenta are of the same sign, 
$L_1L_2>0$. In fact, let $L_1>0$ and $L_2>0$, then $\vp_1$ and $\vp_2$ monotonically grow along the curves and their growth depends only on the corresponding values of $b_1$ and $b_2$. The necessary condition for the intersection at $P_1$ is that there exists $l_1>l_0$ such that $\bar\vp=\vp_1(l_1)=\vp_2(l_1)$. It is clear 
from (\ref{eq:vpl0la}) that difference $\vp_1-\vp_2$ never vanishes if $b_1\neq b_2$.

If $b_1=b_2\equiv b$, then one has $\vp_1(l)=\vp_2(l)$ for all $l$ and the necessary condition holds 
trivially. The sufficient condition of intersection is that for some $l_1$ the time coordinates of $G_1$ and 
$G_2$ are the same, 
$t_1=t_0+k_1 s_1=t_0+k_2 s_2$. Then (\ref{eq:skLl}) implies
	\begin{equation}
\frac{a k_1}{\sqrt{k_1^2-1}}J(b,x_0,x_1)=\frac{a k_2}{\sqrt{k_2^2-1}}J(b,x_0,x_1), 
	\label{eq:vpl0l}
	\end{equation}
or $k_1(k_1-1)^{-\frac{1}{2}}=k_2(k_2-1)^{-\frac12}$ and the condition is satisfied by $k_1=k_2=k$. Finally 
$b_1^2=b_2^2$ implies $L_1^2=L_2^2$ and $L_1=L_2$ --- the geodesics $G_1$ and $G_2$ are identical. 

One infers that $G_1$ and $G_2$ can intersect only if $\vp_1(l)$ grows monotonically $(L_1>0)$ and $\vp_2(l)$ decreases ($L_2<0$). At the intersection point the difference between $\vp_1(l_1)>0$ and $\vp_2(l_1)<0$ is $2\pi$. Applying (\ref{eq:vpl0la}) one gets the necessary condition
	\begin{equation}
\sqrt{1-b_1^2}N(b_1,x_0,x_1)+\sqrt{1-b_2^2}N(b_2,x_0,x_1)=2\pi.
	\label{eq:NN}
	\end{equation}
This is an algebraic equation for $x_1=\frac{l_1}{a}$ at known values of $x_0$, $b_1$ and $b_2$. If a 
solution exists, then the sufficient condition, equality of the time coordinates of $G_1$ and $G_2$, 
$k_1 s_1=k_2 s_2$, requires the following equality to hold, being a direct generalization of (\ref{eq:vpl0l}),
	\begin{equation}
\frac{k_1}{\sqrt{k_1^2-1}}J(b_1,x_0,x_1)=\frac{k_2}{\sqrt{k_2^2-1}}J(b_2,x_0,x_1).
	\label{eq:Jb1Jb2}
	\end{equation}
For $x_1=x_1(x_0,b_1,b_2)$ this is a restriction on the geodesic parameters. Solving both (\ref{eq:NN}) and 
(\ref{eq:Jb1Jb2}) is a hard task, fortunately for our purposes it is sufficient to study a special case of 
these equations. 

The Morris--Thorne wormhole is a globally hyperbolic spacetime and one may apply Theorem 9.12 in~\cite{BEE}.

{\it Theorem}. If $q=G(s_0)$ is the future cut point of $p=G(0)$ along the timelike geodesic $G$ from $p$ to 
$q$, then either one or possibly both of the following hold:
\subitem{(i)} the point $q$ is the first future conjugate point to $p$;
\subitem{(ii)} there exist at least two future directed globally maximal geodesic segments from $p$ to $q$.

By definition, at a conjugate point a non-zero Jacobi vector field along the geodesic vanishes. Jacobi fields 
are solutions to the approximate (linear) geodesic deviation equation hence any Jacobi field connects 
two infinitesimally close geodesic lines. If $G(k,L)$ is a fiducial geodesic surrounded by bundle of close 
geodesics determined by Jacobi fields on $G$, one should distinguish between geodesics lying in the 2-plane 
of $G$ and those directed off the plane. The latter geodesics require a separate treatment given in sect. 3.3 
below and here we comment on close geodesics with $\theta(s)=\pi/2$. 
From the above formulae and discussion one sees that in the wormhole spacetime 
two infinitesimally close geodesics have their angular momenta of the same sign and their parameters $b_1$ and 
$b_2$ must be close, $b_2=b_1+\varepsilon$, $|\varepsilon|\ll 1$, what implies that they will never intersect. 

It is then clear that in the case of ,,coplanar'' curves only distant (besides the end points) geodesics 
can intersect twice. Suppose that $P_1$ 
is the first future cut point to $P_0$ along $G_1$ and let $G_2$ be another globally maximal geodesic from $P_0$ 
to $P_1$ according to the theorem. Their lengths are equal, $s_1=s_2$. At $P_1$ their time coordinates are 
the same, $t_1=t_0+k_1 s_1=t_0+k_2 s_2$, hence their energies are equal, $k_1=k_2=k$. Their parameters $p_1$ and 
$p_2$ satisfy $p_1^2 L_1 ^2=k^2-1=p_2 ^2 L_2 ^2$ and from (\ref{eq:skLl}) it follows
	\begin{equation}
s_1-s_2=0=\frac{a}{\sqrt{k^2-1}} \int\limits_{x_0}^{x_1} \sqrt{x^2+1}
\left[
\frac{1}{\sqrt{x^2+b_1}}- \frac{1}{\sqrt{x^2+b_2}}
\right]\d x.
	\label{eq:s1s2}
	\end{equation}
For $b_1\neq b_2$ the integrand is always either positive or negative and the integral cannot vanish. Hence 
$b_1=b_2$ and this implies $L_1^2=L_2^2$ and then it follows that $L_2=-L_1$. 

One concludes that only two coplanar geodesics, $G_1(k,L)$ and $G_2(k,-L)$ may be both globally maximal 
between points $P_0$ and $P_1$. (Other geodesics emanating from $P_0$ lie in other 2-planes and these planes 
arise due to rotations of the $\theta=\frac{\pi}{2}$ ``plane'' of $G_1$ and $G_2$; in this way the number of 
globally maximal pairs grows to infinity.) 
We remark that $P_1$ is the first cut point of $P_0$ in the sense that $x_1=\frac{l_1}{a}$ is 
the closest to $x_0$ root of eq.~(\ref{eq:NN}). Furthermore, it is clear that $G_1$ and $G_2$ are globally 
maximal between $P_0$ and $P_1$, or that their length $s=d(P_0,P_1)$. In fact, suppose on the contrary, that 
there exists a timelike geodesic $G_3(k_3,L_3)$ from $P_0$ to $P_1$ which is longer, $s(G_3)>s$. This implies 
$b_3\neq b$. Let $L_3 L>0$. By assumption $G_3$ intersects $G_1$ at $P_1$ and this, as was shown above, 
is impossible. $G_3$ does not exist. We emphasize that $G_3$ is not excluded on the assumption that it is 
longer than $G_1$ and $G_2$, also $G_3$ shorter than these two cannot exist. Only two geodesics moving in the opposite directions in coordinate $\vp$ may intersect (modulo rotations of the ``plane''). 

In conclusion, the first future cut point to $P_0$ on $G(k,L)$ is its first intersection point with $G(k,-L)$. 
In this case the necessary condition (\ref{eq:NN}) is simplified to 
	\begin{equation}
\sqrt{1-b^2}N(b,x_0,x_1)=\pi,
	\label{eq:Nbx0x1}
	\end{equation}
showing that the azimuthal angle increases by $\Delta\varphi=\pi$. Then the sufficient condition (40)	
trivially holds.

The indefinite integral $N$ in (37) is the incomplete elliptic integral of the first kind denoted as 
$\F(x,k^2)$, hence eq.~(\ref{eq:Nbx0x1}) reads
	\begin{equation}
\F\left[
\arctan\left(\frac{x_1}{b}\right), 1-b^2
\right]- 
\F\left[
\arctan\left(\frac{x_0}{b}\right), 1-b^2
\right]=\frac{\pi}{\sqrt{1-b^2}}.
	\label{eq:FF}
	\end{equation}
This equation for $x_1=x_1(b,x_0)$ has an exact analytic solution in terms of the Jacobi elliptic function 
$\sc(x,k^2)$,
	\begin{equation}
x_1=b\,\sc\left[
\frac{\pi}{\sqrt{1-b^2}}+\F\left(
\arctan\left(
\frac{x_0}{b}
\right),1-b^2
\right)
\right].
	\label{eq:x1bsc}
	\end{equation}
There are two cases.
	
1. First we consider the special case of timelike geodesics emanating from $P_0$ at the wormhole throat, 
$x_0=0$. In the limit $b\to 0$ the definite integral in (\ref{eq:vpl0la}) is an elementary function which is 
logarithmically divergent in the lower limit $x_0\to 0$. For $x_0=\varepsilon$, $|\varepsilon|\ll1$, one finds 
$x_1\approx |\varepsilon| e^\pi$. For $x_0=0$ and finite $b$ one finds from (\ref{eq:x1bsc}) that $x_1$ 
rapidly grows to infinity for increasing $b$: finite solutions $x_1(b)$ exist only in the narrow interval 
$0<b<b_M=0,16780{\ldots}$ For $b>b_M$ timelike geodesics crossing the throat have no cut points to 
$P_0(x_0=0)$, hence are globally maximal up to $l\to +\infty$. The limit $b<b_M$ corresponds to 
$L^2>0{,}971843(k^2-1)a^2$.

Since the geodesics $G_1(k,L)$ and $G_2(k,-L)$ which intersect at $x_1$ (if this point exists) are complete, 
one may also consider their behaviour for $l\to -\infty$. If the upper limit in the integral (\ref{eq:vpl0la}) 
is $x<0$, then $\vp_1(l)$ monotonically diminishes and $\vp_2(l)$ monotonically grows for $l\to -\infty$. For 
$l<0$ geodesics $G_1$ and $G_2$ will intersect in the first past cut point $P_2(l_2)$, $l_2<0$, if 
$\vp_1(l_2)-\vp_2(l_2)=-2\pi$. The integrand in (\ref{eq:vpl0la}) is a symmetric function since it depends on 
$x^2$, hence the incomplete elliptic integral of the first kind is antysymmetric, $\F(-x,k^2)=-\F(x,k^2)$,  therefore if $x_1(b)$ is the lowest positive solution to (\ref{eq:Nbx0x1}), then $x_2=-x_1$ is the largest 
negative solution of the corresponding equation $\sqrt{1-b^2}\,N(b,0,x_2)=-\pi$.

One summarizes the case $x_0=0$ by stating that if $b(k,L)$ is in the interval $0<b<0{,}16780{\ldots}$, then 
the geodesics $G_1(k,L)$ and $G_2(k,-L)$ intersect at $l=l_1>0$ and at $l=l_2=-l_1$. If $0{,}16780<b<1$, 
then geodesics $G_1$ and $G_2$ are globally maximal from $l=0$ to $l\to \pm\infty$. The explicit form of 
$G_1$ (and correspondingly of $G_2$) is $t=t_0+ks$, $s(k,L,l)=a(k^2-1)^{-\frac12} J(b,x_0=0,\frac{l}{a})$ 
(eq.~(\ref{eq:skLl})), $\theta=\frac\pi 2$ and $\vp(l)$ as in (\ref{eq:vpl0la}). $G_1$ is uniquely 
determined by its tangent vector $\dot{x}^\alpha$ at the initial crossing point $P_0(x_0=0)$,
	\begin{equation}
\dot{x}^\alpha =\left[
k,\left(k^2-1-\frac{L^2}{a^2}\right)^\frac{1}{2}, 0, \frac{L}{a^2}
\right].
	\label{eq:xdotalpha}
	\end{equation}
At $P_0$ the ratio of the components of the tangent vector is
	\begin{equation}
\frac{{\dot l}}{{\dot\vp}}=\frac{a\,b}{\sqrt{1-b^2}}.
	\label{eq:dotlvarphi}
	\end{equation}
It follows that if $b<b_M$ then $\frac{\dot l}{\dot\vp}<a\, b_M(1-b_M^2)^{-\frac12}=0{,}17021{\ldots}$ and 
the geodesics $G_1$ and $G_2$ have cut points to $P_0$ at $l=l_1$ and at $l=-l_1$; 
for $\frac{\dot l}{\dot \vp}>0{,}17021{\ldots}$ there are no cut points to $l_0=0$ and $G_1$ and $G_2$ are 
globally maximal up to $l\to \pm\infty$.

\begin{figure}[h]\vspace*{0.75ex}
		\begin{center}
\hspace*{-1em}\includegraphics[width=0.65\textwidth]{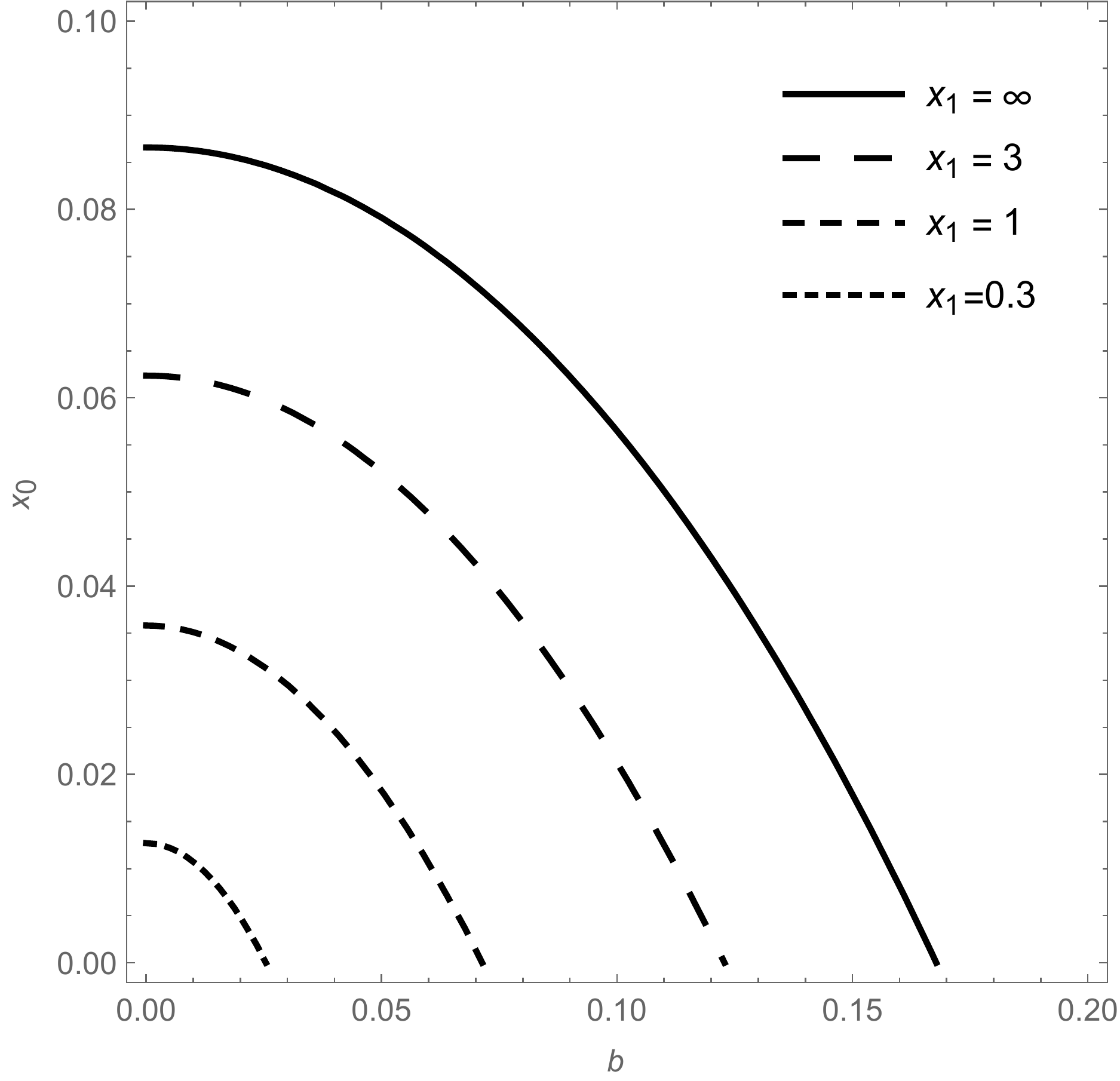}
		\end{center}
\vspace{-4ex}
\caption{Four curves representing values of $b$ and $x_0$ corresponding to three finite values of $x_1$ and 
to $x_1=\infty$. For $b>b_M=0{,}16780{\ldots}$ there are no solutions.}
\label{Figure01}
\vspace{-1mm}
		\end{figure}

2. In the general case geodesics $G_1(k,L)$ and $G_2(k,-L)$ emanate from $P_0$ for $x_0>0$. Again the solution 
$x_1(b,x_0)$ given in (\ref{eq:x1bsc}) does not exist for all $x_0>0$ and $b^2<1$. On the plane $(b,x_0)$ a 
finite solution exists for points in the domain bounded by the coordinate axes and the limiting curve 
corresponding to $x_1=\infty$. The limiting curve is given by exact equation 
	\begin{equation}
-\F\left[\arctan\left(\frac{x_0}{b}\right), 1-b^2\right]+\K(1-b^2)=\frac{\pi}{\sqrt{1-b^2}},
	\label{eq:asymptotyka}
	\end{equation}
where $\K(k^2)$ is the complete elliptic integral of the first kind, see~Figure~\ref{Figure01}.
For small $b$ the root $x_1$ rapidly blows up to infinity for small $x_0$: one finds numerically from 
(\ref{eq:asymptotyka}) that for $b=10^{-7}$ the maximal value of $x_0$ is $x_0=0{,}08662{\ldots}$ Hence in 
the strip $0<b<1$, $0<x_0<\infty$ most of the region corresponds to globally maximal geodesics. 

\subsection{Conjugate points on non-radial timelike geodesics}

Let $G(k,L)$ be a general non-radial timelike geodesic lying in the 2-plane $\theta=\frac{\pi}{2}$. Its 
tangent vector is 
	\begin{equation}
u^\alpha=\frac{\d x^\alpha}{\d s}=\left[
k,\frac{p L}{\sqrt{x^2+1}}\left(
x^2+b^2
\right)^\frac12,0,\frac{L^2}{a^2(x^2+1)}
\right].
	\label{eq:ualpha48}
	\end{equation}
$G$ is viewed as a fiducial geodesic for a bundle of infinitesimally close to it geodesics connected to $G$ 
by various Jacobi fields. These vector fields are solutions of the geodesic deviation equation on $G$ and the 
general formalism for finding these fields and conjugate points determined by them is presented in~\cite{SG3}. 
According to it one introduces a spacelike basis triad along $G$, $e_a{}^\mu(s)$, $a=1,2,3$ and any Jacobi 
vector field is spanned in this basis, $Z^\mu=\sum_a Z_a e_a{}^\mu$, where $Z_a(s)$ are called Jacobi scalars. 
The basis triad for a generic USSS spacetime is given~\cite{SG1} and after adopting it to $(t,l,\theta\vp)$ coordinates it reads 
{\setlength\arraycolsep{2pt}
	\begin{eqnarray}
e_1{}^\mu&=&\left[
0,\frac{1}{p a\sqrt{x^2+1}},0,-\frac{(x^2+b)^\frac12}{a(x^2+1)}
\right],~~e_2{}^\mu=\frac{1}{a\sqrt{x^2+1}}\delta_2^\mu,\nonumber\\
e_3{}^\mu&=&
\left[
\sqrt{k^2-1},\frac{k}{\sqrt{x^2+1}}(x^2+b^2)^\frac12,0,\frac{k}{p a^2(x^2+1)}
\right].
	\label{eq:baza}
	\end{eqnarray}}
Also generic solutions for Jacobi scalars are given in~\cite{SG1}. The scalar $Z_3$ is a linear function 
$Z_3=C_{31}s +C_{32}$ ($C_{31}$ and $C_{32}$ are arbitrary constants) in every USSS spacetime and it is clear 
that the deviation vector $Z^\mu=Z_3 e_3{}^\mu$ does not generate conjugate points on $G$. The scalar $Z_1(s)$ 
is expressed in terms of integrals of the metric functions and for the metric (\ref{eq:metricUSSS}) one gets 
	\begin{equation}
Z_1(l(s))=C_1(x^2+b^2)^\frac12 \left[
\int\!\!\frac{(x^2+1)^\frac12}{(x^2+b^2)^\frac32}\d x-C_0
\right]\equiv C_1(x^2+b^2)^\frac12\left[Y(b,x)-C_0
\right].
	\label{eq:Z1ls}
	\end{equation}
By definition, the Jacobi vector field $Z^\mu=Z_1 e_1{}^\mu$ must vanish at the initial point $P_0$, 
$x_0=\frac{l_0}{a}$, and this determines the constant $C_0$ as $C_0=Y(b,x_0)$. Then there exists a conjugate point 
$P_1$ to $P_0$ for $x_1\neq x_0$ if $Z^\mu(x_1)=0$ or $Z_1(x_1)=0$. However $Y(b,x)$ is a monotonically  
increasing function and $Y(b,x_1)=C_0$ implies $x_1=x_0$. Thus one sees that also the geodesic generated from $G$ by means of the field $Z_1 e_1{}^\mu$ (actually the factor $C_1$ in (\ref{eq:Z1ls}) shows that there is a one-dimensional family of geodesics determined by $Z_1 e_1{}^\mu$) may intersect $G$ only once at $P_0$. The non-existence of intersection points of $G$ with the geodesics generated by Jacobi vectors $Z_1 e_1{}^\mu$ and 
$Z_3 e_3{}^\mu$ is in full agreement with the result shown in sec.~3.2. In fact, the basis vectors $e_1{}^\mu$ 
and $e_3{}^\mu$ have their $\theta$-components equal zero, what means that the geodesics generated by the corresponding Jacobi vector fields also lie in $\theta=\frac{\pi}{2}$ plane, while we have shown above that 
among coplanar geodesics only $G(k,L)$ and $G(k,-L)$ may intersect twice. Therefore the most interesting 
Jacobi field is that $Z_2 e_2{}^\mu$ which goes off the $\theta=\frac{\pi}{2}$ plane. It reads 
	\begin{equation}
Z_2 e_2{}^\mu=\left(
C_{21}\cos\vp(l)+C_{22}\sin\vp(l)
\right)\delta_2^\mu .
	\label{eq:Z2e2}
	\end{equation}
Assuming that the initial point $P_0$ has $\vp_0=0$ one considers the field 
$Z_\mu=C_2\delta_2^\mu \sin\vp(l)$ and possible conjugate points to $P_0$ are $\vp_n(l)=n\pi$, $n=1,2{\ldots},$. 
The angle $\vp$ on $G$ is given in (\ref{eq:vpl0la}) and the first conjugate point is for $\vp(l_1)=\pm\pi$ or 
$\sqrt{1-b^2}N(b,x_0,x_1)=\pi$. This is exactly the point $P_1$ of intersection of geodesics $G(k,L)$ and 
$G(k,-L)$ determined by eq.~(\ref{eq:Nbx0x1}). It follows from the previous considerations that this is the 
first and a single cut points to $P_0$ on the geodesic $G(k,L)$. (Each conjugate point is a cut point, but in general cut points are not conjugate ones.)

\section{Conclusions}

Our search for the cut points on timelike geodesic curves shows a rather complicated structure even in a very 
simple case of an ultrastatic spherically symmetric spacetime such as Morris--Thorne wormhole with the metric 
(\ref{eq:metricUSSS}). A generic timelike geodesic is  described by elliptic integrals $\F(x,k^2)$ and 
$\K(k^2)$ and its future (and past) cut points are determined by these functions. Cut points are intersection 
points of coplanar ($\theta=\pi/2$) geodesics of equal energy and equal and oppositely directed angular momenta. 
These points are also conjugate points determined by infinitesimally close geodesics lying in different 2-planes 
which are close to each other and intersecting. If a fiducial geodesic has a cut point to a given initial 
point on it, it is intersected there by infinite number of other geodesics emanating from the initial point.  
The 
cut points exist only for energies and angular momenta belonging to very narrow intervals; otherwise timelike geodesics are globally maximal on their complete segments. This feature cannot be recognized directly from 
the metric (or curvature), one must apply the properties of the elliptic integrals to establish it. 

One may compare Morris--Thorne wormhole with another USSS spacetime, the global Barriola--Vilenkin 
monopole~\cite{BV,Ch}. The comparison is incomplete since in the latter spacetime only conjugate points on 
non-radial geodesics are known \cite{SG1}. In both the spacetimes the conjugate points are determined by 
the Jacobi field directed off the $\theta=\pi/2$ plane (the field is proportional to the vector 
$e_2{}^{\mu}$ of the spacelike basis along the geodesic), hence coplanar close geodesics cannot intersect 
twice. In the monopole spacetime the conditions for the existence of conjugate points are determined by 
the parameter appearing in its metric. Yet the wormhole metric is parameter--free and conditions for cut 
points to exist are deeply hidden in the spacetime geometry. 

\section*{Appendix. Ultrastatic spacetimes and inertial reference frames}

\setcounter{equation}{0}
\renewcommand{\theequation}{A.\arabic{equation}}

In most textbooks on classical mechanics (see e.g.~\cite{LL1,GPS}) and also in many ones on special relativity 
(e.g.~\cite{LL2}) it is claimed that the inertial reference frame may be uniquely defined in purely 
dynamical terms: as that frame in which a freely moving body (i.e. one which is not acted upon by external 
forces) moves uniformly with constant velocity or remains at rest. Discovery of ultrastatic spacetimes has 
shown that this definition is not unique since any coordinate system in these spacetimees which explicitly 
exhibits that they are ultrastatic, dynamically imitates the inertial frame. It has been found that in 
these spacetimes there are no ``true'' gravitational forces, only ``inertial'' ones. More precisely, 
ultrastatic spacetimes are defined as those which admit a timelike Killing vector field which is covariantly constant $\nabla_\alpha K_\beta=0$. This implies that the field does not accelerate (expand), rotate or deform 
and the reference frame determined by this vector field mimicks  the notion of the inertial frame in 
Minkowski spacetime (see~\cite{SS} and numerous references on inertial forces therein). In ultrastatic 
spherically symmetric spacetimes it was shown that a free particle has a constant velocity relative to the 
comoving frame (its motion is uniform)~\cite{SG1}.

Below we prove a generic theorem showing to what extent a comoving coordinate system in a generic ultrastatic spacetime imitates the true inertial frame. From its definition, the timelike Killing vector characterizing 
any ultrastatic spacetime is a gradient, $K_\alpha=\partial_\alpha t$ and has a constant norm. Then it may 
be chosen as $K_\alpha=\delta_0^\alpha$ and the corresponding system is comoving (Gauss normal geodesic) with 
the metric~\cite{SS}
	\begin{equation}
\d s^2=\d t^2+g_{ik}(x^j)\d x^i\d x^k,~~i,j,k=1,2,3
	\label{eq:A1}
	\end{equation}
where $\det(g_{ik})<0$. Distinct ultrastatic spacetimes differ in the three-metric $g_{ik}$, which is time-independent. 

{\it Proposition}. A timelike geodesic motion in any ultrastatic spacetime may be described as a free motion in 
the constant time 3-space subject to covariant nonrelativistic Newtonian equations of motion with vanishing 
force. The free motion has following properties:
\litem{i))} the 3-velocity has a constant norm, hence the motion is uniform; 
\litem{ii)}its trajectory is a geodesic of the 3-space. 

{\it Proof}. The timelike Killing vector $K^\alpha$ generates along any timelike geodesic the integral of 
energy per unit mass $k=\frac{E}{m\,c^2}$ and $\frac{\d t}{\d s}=k>0$ from~(\ref{eq:geodesict}). The 
$t=\mbox{const}$ space has the metric
	\begin{equation}
\d \sigma^2=\gamma_{ij}(x^k)\d x^i\d x^j
	\label{eq:A2}
	\end{equation}
where $\gamma_{ij}=-g_{ij}$ (the signature is $+\,-\,-\,-$). The metric $\gamma_{ij}$ determines the connection 
$\overline{\mathit\Gamma}_{jk}^i(\gamma)$. The spacetime connection is  ${\mathit\Gamma}_{jk}^i(g)=
\overline{\mathit\Gamma}_{jk}^i(\gamma)$, other components of ${\mathit\Gamma}_{\mu\nu}^\alpha(g)$ vanish. 
The timelike geodesic equation has three independent components, 
	\begin{equation}
\frac{\rm D}{\d s}\frac{\d x^i}{\d s}=\frac{\d^2 x^i}{\d s^2}+ {\mathit\Gamma}_{jk}^i(g)\frac{\d x^j}{\d s}
\frac{\d x^k}{\d s}=0,
	\label{eq:A3}
	\end{equation}
whereas $\frac{\rm D}{\d s}\frac{\d x^0}{\d s}=\frac{\d}{\d s}k\equiv 0$. The connection 
$\overline{\mathit\Gamma}_{jk}^i(\gamma)$ in the space generates two absolute derivatives with respect to 
the scalars $\sigma$ and $t$, $\frac{\overline{\rm D}}{\d s}$ and $\frac{\overline{\rm D}}{\d t}$. One defines 
the particle's 
3-velocity $v^i\equiv\frac{\d x^i}{\d t}$ along the geodesic, which is a 3-vector with respect to purely spatial coordinate transformations. From $\frac{\d t}{\d s}=k$ one gets $v^i=\frac{1}{k}\frac{\d x^i}{\d s}$. 

Next one postulates, as in classical mechanics, that any particle motion in the space is subject to 
nonrelativistic Newtonian equations of motion, 
	\begin{equation}
\frac{\overline{\rm D}}{\d t}\left(m\,v^i\right)\equiv m
\left[
\frac{\d v^i}{\d t}+ \overline{\mathit\Gamma}_{jk}^i v^j v^k
\right]
=F^i(t,x^j),
	\label{eq:A4}
	\end{equation}
where $F^i$ is some external force. In the case of a free particle (no other interactions besides gravitation) 
one has $\frac{\d}{\d s}\frac{\d x^i}{\d s}=k^2\frac{\d v^i}{\d t}$ and the three geodesic equations 
(\ref{eq:A3}) take on the form
	\begin{equation}
\frac{{\rm D}}{\d s}\frac{\d x^i}{\d s}
\equiv k^2\left[
\frac{\d v^i}{\d t}+ \overline{\mathit\Gamma}_{jk}^i v^j v^k
\right]=0
	\label{eq:A5}
	\end{equation}
implying that the force in the Newtonian equations~(\ref{eq:A4}) vanish, $F^i=0$. Hence the free (geodesic) 
motion in the ultrastatic spacetime also manifests itself as a free motion in the space (if parametrized by 
time $t$ as an external parameter).

The length $V$ of the 3-velocity is $V^2=\gamma_{ij}v^i v^j>0$ and (\ref{eq:A4}) shows that it is constant,  			\begin{equation}
\frac{\d}{\d t}V^2=\frac{\overline{\rm D}}{\d t}\left(\gamma_{ij} v^i v^j\right)=0,\nonumber 
	\label{eq:A6}
	\end{equation}
or the motion in the space is uniform. Clearly this motion is not rectilinear since straight lines in general 
do not exist in curved spaces, yet the trajectory $x^i=x^i(t)$ is a geodesic of the space. In fact, along the trajectory in the space one has $\d x^i=v^i\d t$ and $\d \sigma^2=V^2\d t^2$, hence $\d\sigma=V\d t$ and the relationship is linear, $\sigma=V\, t+\mbox{const}$. The trajectory is parameterized by its length, 
$x^i=x^i\left(t(\sigma)\right)$. Its tangent vector $\frac{\d x^i}{\d\sigma}=\frac{v^i}{V}$ satisfies 
	\begin{equation}
\frac{\overline{\rm D}}{\d \sigma}\frac{\d x^i}{\d\sigma}=
\frac{\d t}{\d \sigma}
\frac{\overline{\rm D}}{\d t}
\left(
\frac{v^i}{V}
\right)=\frac{1}{V^2}\frac{\overline{\rm D}}{\d t}v^i=0;\nonumber 
	\label{eq:A7}
	\end{equation}
the trajectory in the space generated by the timelike geodesic in the spacetime is a geodesic of this space. 

If an ultrastatic spacetime is also spherically symmetric (USSS), then radial timelike geodesics perfectly imitate 
the inertial motion in Minkowski space since they are straight lines on Minkowski 2-plane. In fact, in the 
adapted coordinates the metric reads
\begin{equation}
\d s^2= \d t^2- e^{\lambda(r)} \d r^2- r^2 \d {\mathit\Omega}^2,   
	\label{eq:A8}
	\end{equation}
where $r\in(r_m,r_M)$. One introduces a new radial coordinate $\rho$ by $\d r=e^{-\lambda/2} \d\rho$, then 
\begin{equation}
\rho=\int\!\! e^{\lambda/2} \d r\equiv W(r).
	\label{eq:A9}
	\end{equation}
Function $W(r)$ is invertible in the whole range of $r$ and $r=W^{-1}(\rho)$. The metric is now 
\begin{equation}
\d s^2= \d t^2- \d \rho^2- r^2(\rho) \d {\mathit\Omega}^2.   
	\label{eq:A10}
	\end{equation}		
A radial geodesic C is $t=t(s)$, $r=r(s)$, $\theta=\theta_0$ and $\varphi=\varphi_0$. In the coordinates 
$(t,\rho,\theta,\varphi)$ its tangent vector is $\dot{x}^{\alpha}=(\dot{t},\dot{\rho},0,0)$ where 
$\dot{t}=k$. Then the normalization $g_{\alpha\beta} \dot{x}^{\alpha} \dot{x}^{\beta}=1$ yields 
$\dot{\rho}=+\sqrt{k^2-1}$ for outgoing C and finally $t=t_0+ks$ and $\rho(s)=\sqrt{k^2-1} s+\rho_0$. 
C is a straight line on Minkowski plane $(t,\rho)$. By a hyperbolic rotation on the plane each radial C 
may be identified with a time coordinate line on this plane. (However this hyperbolic rotation is not an 
ultrastatic transformation according to the definition given in \cite{SS}.)

The genuine inertial frame exists only in Minkowski spacetime and should be defined in purely geometric terms. 

\section*{Acknowledgments}

The work of both the authors was supported by the John Templeton Foundation Grant ``Conceptual Problems in Unification Theories" no. 60671.

\end{document}